\documentclass[showpacs,eqsecnum,aps]{revtex4}
\usepackage{amsmath}
\usepackage{amssymb}
\usepackage{amsfonts}
\usepackage{graphicx}
\def\beq{\begin{equation}}
\def\eeq{\end{equation}}
\def\beqa{\begin{eqnarray*}}
\def\eeqa{\end{eqnarray*}}

\def\bLam{\mbox{\boldmath$\Lambda$}}

\begin{document}

\title{Cluster properties and particle production in
Poincar\'e invariant quantum mechanics}

\author{W. N. Polyzou\footnote{This work supported 
in part by the U.S. Department of Energy, under contract DE-FG02-86ER40286}}
\affiliation{
Department of Physics and Astronomy, The University of Iowa, Iowa City, IA
52242}

\vspace{10mm}
\date{\today}

\begin{abstract}
I outline the construction of exactly Poincar\'e invariant
quantum models that satisfy cluster separability but do not conserve particle 
number.
\end{abstract}
\vspace{10mm}

\maketitle

\section{Introduction}
%\label{SchmidtPL_intro}
\label{PolyzouWP_intro}
%Your text comes here ...
%The \LaTeX~source file of this document constitutes a template for the style
%and layout of the proceedings of the the 19$^{\textnormal{\footnotesize th}}$
%International IUPAP Conference on Few-Body Problems in Physics (FB19), held
%from August 31 to September 5, 2009 in Bonn, Germany.
%xxxxxxxxxxxxxxxxxxxxxxxxxxxxxxxxxxxxxxxxxxxxxxxxxxxxxx

Few-body physics has been transformational in terms of how we
understand low-energy nuclear physics.  The success is largely due to
the existence of uncoupled few-body problems that are directly related
to experiment and cluster properties that relate the few and many-body
Hamiltonians.  Thus, nucleon-nucleon interactions are fine tuned by
comparing cross sections calculated using numerically exact solutions
of the Lippmann-Schwinger equation to experimental nucleon-nucleon
cross sections.  Cluster properties fix how these nucleon-nucleon
interactions appear in N$>$2-body Hamiltonians.  Small corrections due
to three-body interactions can also be fine tuned by comparing
numerically exact solutions of the Faddeev equations with experiment.
Cluster properties again fix how the two and three-body interactions
appear in the N$>$3-body Hamiltonians.  The saturation of nuclear
binding energies suggests that at nuclear densities two, three and
possibly four-body interactions are sufficient to construct
Hamiltonians that provide an accurate description of most nuclei.  The
level of success achieved in low-energy nuclear physics has not been
duplicated for energy scales above the threshold for the production of
pions.

There are a number of reasons for the increased difficulty:

1.) For energies approaching or above the GeV scale the Poincar\'e
group must be a symmetry of the theory so calculations in the
laboratory and center of momentum frame are consistent.  Cluster
properties become more difficult to satisfy in Poincar\'e invariant
models.  While cluster properties can be satisfied for fixed number of
particles\cite{sokolov}\cite{fcwp82}, there is no known systematic
treatment of cluster properties for few-body systems where particle
number is not conserved.

2.) A consistent treatment of particle production also requires a
Poincar\'e symmetric treatment.  Particle production violates Galilean
invariance; momentum conservation cannot be simultaneously satisfied
in two frames related by Galilean boosts.

3.) There is no few-body problem directly related to experiment.  Even
the one-nucleon problem involves an infinite number of bare pions.
This makes it difficult to construct phenomenological few-body
interactions by comparing numerically exact calculations to
experiment.

4.) Even if one can satisfy cluster properties, asymptotically
separated clusters still involve infinite numbers of bare particle
degrees of freedom.

The few-GeV scale is an important energy scale.  It is the scale where
it is possible to study the sensitivity of the dynamics to sub-nucleon
degrees of freedom.  Also, even though the physics involves an
infinite number of bare particle degrees of freedom, we expect that
the dynamics should be dominated by a finite number of suitably chosen
degrees of freedom; it is hard to believe that a small change in
energy would require the number of degrees of freedom needed to
describe nucleon-nucleon scattering to suddenly jump from two to
infinity.

In this talk I discuss a strategy to address all of the difficulties
discussed above, extending the few-body program so it can be used to
make realistic models of systems with invariant energy above the
pion-production threshold that preserve the properties that have made
few-body methods transformational below the pion-production threshold.
The physics motivation for doing this is to extend the success of
few-body methods to energy scales where sub-nucleon degrees of freedom
become relevant.
This is also relevant because it is now possible to solve relativistic 
Faddeev equations at the GeV scale \cite{ce1}\cite{ce2}\cite{ce3}. 
 
To treat this problem it is useful to consider the structure of
low-energy few-body models.  In principle, from the point of view of
perturbative field theory, even low-energy nucleon-nucleon scattering
involves an infinite number of bare meson degrees of freedom.
Physical nucleons can also be understood as bare nucleons with a cloud
of virtual mesons.  However, in low-energy nuclear physics one never
deals with the bare nucleon degrees of freedom; instead the relevant
degrees of freedom are taken as physical nucleons.  Their
masses are not calculated, they are measured.  Even though the
structure of the nucleons cannot be calculated, it can still be probed
using scattering experiments.  For example, electron proton-scattering
suggests that the proton has a non-trivial electromagnetic structure.
The fundamental vertex involving bare mesons and nucleons is replaced
by a phenomenological nucleon-nucleon interaction between physical
nucleons.  While the Lippmann-Schwinger equation with realistic
nucleon-nucleon interactions can be solved at all energies, it is only
designed to be taken seriously for energies where the dominant degrees
of freedom are two physical nucleons.  This means that the number of
relevant degrees of freedom is dictated by the energy scale of
interest.  All of this can even be done relativistically using
Poincar\'e symmetric versions of the two-nucleon problem.

The observations in the previous paragraph suggest how to proceed.
Building on the what was successful in the low-energy case, we
formulate the theory using only physical-particle degrees of freedom.
Thus, nucleons, deuterons, and pions will all be taken as physical
particles.  Since physical particles have no-self interactions, a
theory formulated in terms of physical particle degrees of freedom
cannot have elementary vertices, otherwise there would be mass
renormalizations.  Interactions must be short-ranged, involving two or
more initial and two or more final particles.  The simplest
interaction that changes particle number is a short-ranged $2
\leftrightarrow 3$ interaction.  Even if the theory is formulated
using physical particle degrees of freedom, if there is real particle
production, the theory still involves an infinite number of degrees of
freedom.  To have few-body problems directly constrained by experiment
it is necessary to replace a single theory that is applicable at all
energy scales by an ordered sequence of effective theories with
successively more degrees of freedom that are relevant over different
energy scales.  When a system of physical particles is broken up into
asymptotically separated subsystems, the subsystems will involve
lower-energy scales.  Cluster properties can be realized if these
lower energy sub-systems are constrained by the lower-energy effective
theories.  
%In this approach there is no nuclear structure problem,
%however nuclear structure is actually probed using scattering
%experiments, which can be formulated in this framework.

In what follows I use the example of nucleon-nucleon scattering 
between the one and two pion-production threshold to illustrate the
structure and construction of such a theory. 

\section{Hilbert space}
\label{WP_sec:2} 

Minimally, the model Hilbert space must have enough structure so that
it is possible to compute probability amplitudes for any experiment of
interest.  For experiments at a fixed energy scale, it is possible to
measure the momenta and spin projections of every particle that can
appear in an initial or final state.  This motivates the choice of
model Hilbert space as the direct sum of tensor products of single
physical-particle Hilbert spaces, where the particle content
corresponds to the particles that can be experimentally observed at
the given energy scale.  This space has enough degrees of freedom to
describe the results of any experiment that can be performed at the
given energy scale.

Complete experiments on isolated physical particles measure their
mass, linear momentum,  spin, and spin polarization with respect to
some axis in a given frame.  A suitable representation of the
one-particle Hilbert space, ${\cal H}_1$, is the space of square
integrable functions, $\psi (\mathbf{p}, \mu )$, of these observables
\beq
\psi (\mathbf{p}, \mu ) =
\langle (m,j) \mathbf{p}, \mu
\vert \psi \rangle 
\label{a.1}
\eeq
\beq
\langle \psi \vert \psi \rangle =
\int 
\sum_{\mu=-j}^j\vert \psi (\mathbf{p}, \mu ) \vert^2 d\mathbf{p} < 
\infty 
\label{a.2}
\eeq
where $(m,j)$ are the physical mass and spin of the particle.

The formulation of cluster properties works best when the theory is
formulated in terms of tensor products.  In order to preserve the
tensor product structure in models with particle production, it is
useful to replace the single particle Hilbert space by the direct sum
of a single particle Hilbert space and a zero dimensional no-particle
space:
\beq
{\cal H}_1 \to {\cal H}_1 \oplus \{0 \} .
\label{a.3}
\eeq
These doublet spaces were first introduced by Sokolov\cite{sokolov}.  
The tensor product of doublet spaces can be decomposed as a direct sum of
tensor products.  For example, if the
system has particles of type $b$,$c$,and $d$, the tensor product of
the three doublet spaces can be expanded into a direct sum of
subspaces with all particle contents involving particles $b$,
$c$, and $d$.
\[
({\cal H}_b\oplus \{0\}) \otimes 
({\cal H}_c\oplus \{0\}) \otimes 
({\cal H}_d\oplus \{0\}) 
 =
\]
\[
({\cal H}_b \otimes 
{\cal H}_c \otimes 
{\cal H}_d ) \oplus
({\cal H}_b \otimes 
{\cal H}_c ) 
\oplus 
({\cal H}_c \otimes 
{\cal H}_d ) 
\oplus
\]
\beq
({\cal H}_d \otimes 
{\cal H}_b ) 
\oplus 
{\cal H}_b\oplus  
{\cal H}_c\oplus 
{\cal H}_d\oplus \{0\} .
\label{a.4}
\eeq 
In specific models some of these combinations do not appear because of
super-selection rules, energy considerations, or properties of the
interactions.  In what follows, we construct our models
assuming that the particles are initially distinguishible, and then
treat the particle identity when computing the cross sections.

The single-particle spaces are also irreducible representation spaces
for the Poincar\'e group.  The unitary irreducible representation,
$U_1(\Lambda ,a)$, of the Poincar\'e group on the single particle 
Hilbert space ${\cal H}_1$ is
\[
\langle (m,j) \mathbf{p}, \mu \vert U_1(\Lambda ,a)
\vert \psi \rangle = 
\]
\[
\int  \sum_{\mu'=-j}^j 
\langle (m,j) \mathbf{p}, \mu \vert U_1(\Lambda ,a)
\vert (m,j) \mathbf{p}', \mu' \rangle d\mathbf{p}'
\times
\]
\[
\langle (m,j) \mathbf{p}', \mu'
\vert \psi \rangle = 
\]
\beq
\int  \sum_{\mu'=-j}^j 
{\cal D}^{mj}_{\mathbf{p}, \mu; \mathbf{p}', \mu'}
[\Lambda ,a] d\mathbf{p}' \psi (\mathbf{p}', \mu' )
\label{a.5}
\eeq
where the Poincar\'e group Wigner function is 
\[
{\cal D}^{mj}_{\mathbf{p}, \mu; \mathbf{p}', \mu'}
[\Lambda ,a] :=
\langle (m,j) \mathbf{p}, \mu \vert U_1(\Lambda ,a)
\vert (m,j) \mathbf{p}', \mu' \rangle =
\]
\beq
e^{i p \cdot a}
\sqrt{{\omega_m (\mathbf{p}) \over \omega_m (\mathbf{p}') }} 
\delta (\mathbf{p} - \bLam p' ) 
D^i_{\mu \mu'}[B^{-1} (p/m)\Lambda B(p',m) ]  
\label{a.6}
\eeq
with
\beq
\omega_m (\mathbf{p}) = \sqrt{\mathbf{p}^2 + m^2} 
\label{a.7}
\eeq
and $B(p/m)$ is a Lorentz boost 
\beq 
B(p/m) (m,0,0,0) = p 
\label{a.8}
\eeq
that depends on the choice of spin polarization observable (helicity,
canonical spin, null-plane spin).  The quantity 
$B^{-1} (p/m)\Lambda B(p'/m)$ 
is a Wigner rotation.  

%The Poincar\'e group Wigner function
%${\cal D}^{mj}_{\mathbf{p}, \mu; \mathbf{p}', \mu'}
%[\Lambda ,a]$ satisfies the 
%group representation property
%\beq
%\int d\mathbf{p}'' \sum_{\mu''=-j}^j 
%{\cal D}^{mj}_{\mathbf{p}, \mu; \mathbf{p}'', \mu''}
%[\Lambda_2 ,a_2] \,
%{\cal D}^{mj}_{\mathbf{p}'', \mu''; \mathbf{p}', \mu'}[\Lambda_1 ,a_1]
%={\cal D}^{mj}_{\mathbf{p}, \mu; \mathbf{p}', \mu'}
%[\Lambda_2 \Lambda_1, \Lambda_2 a_1 + a_2].
%\eeq
The direct sum of tensor products of these single-particle
unitary irreducible representations, where the representation acts 
like the identity on the zero dimensional no-particle state,
defines the non-interacting representation, $U_0(\Lambda,a)$, of 
the Poincar\'e group
on the model Hilbert space.  On the tensor product of the three
doublet spaces (\ref{a.4}) it has the form
\[
U_0(\Lambda ,a) = 
[U_a(\Lambda ,a) \otimes 
U_b(\Lambda ,a) \otimes 
U_c(\Lambda ,a) ] \oplus
\]
\[
[U_a(\Lambda ,a) \otimes 
U_b(\Lambda ,a) ] \oplus
[U_b(\Lambda ,a) \otimes 
U_c(\Lambda ,a) ] \oplus
\]
\[
[U_c(\Lambda ,a) \otimes 
U_a(\Lambda ,a) ] \oplus
U_a(\Lambda ,a) \oplus 
\]
\beq
U_b(\Lambda ,a) \oplus 
U_c(\Lambda ,a)  \oplus I .
\label{a.9}
\eeq

\section{Nucleon-nucleon scattering model}
\label{WP_sec:3} 

Consider nucleon-nucleon scattering for invariant energy between
$2m_N + m_\pi$ and $2m_N + 2m_\pi$.  In this energy range
the possible baryon number two states are 
$(NN)$, $(NN\pi)$, $(D\pi)$, $(D\pi\pi)$ .
Using Sokolov's doublet formalism and treating the nucleons and
pions as distinguishible, the model Hilbert space is the 
following direct sum of 
tensor products of two and three particle spaces: 
\[
{\cal H} = {\cal H}_{N_1N_2} \oplus 
{\cal H}_{N_1N_2\pi_1} \oplus
{\cal H}_{N_1N_2\pi_2} \oplus
\]
\beq
{\cal H}_{D\pi_1} \oplus
{\cal H}_{D\pi_2} \oplus
{\cal H}_{D\pi_1\pi_2}.
\label{a.10}
\eeq
Vectors in this space are represented by six-component wave functions:
\beq
\Psi (\cdots )=
\left (
\begin{array}{c}
\psi_{NN} (\mathbf{p}_N,\mu_N,\mathbf{p}_{N'},\mu_{N'} )\\
\psi_{D\pi} (\mathbf{p}_D,\mu_D,\mathbf{p}_{\pi}  )\\
\psi_{D\pi'} (\mathbf{p}_D,\mu_D,\mathbf{p}_{\pi'}  )\\
\psi_{NN\pi} (\mathbf{p}_N,\mu_N,\mathbf{p}_{N'},\mu_{N'},\mathbf{p}_{\pi}  )\\
\psi_{NN\pi'} (\mathbf{p}_N,\mu_N,\mathbf{p}_{N'},\mu_{N'},\mathbf{p}_{\pi'}  )\\
\psi_{D\pi\pi} (\mathbf{p}_D,\mu_P,\mathbf{p}_{\pi},\mathbf{p}_{\pi'})\\
\end{array}
\right ) .
\label{a.11}
\eeq

The kinematic unitary representation of the Poincar\'e group is
\[
U_0(\Lambda ,a) = 
[U_{N}(\Lambda ,a) \otimes U_{N'}(\Lambda ,a)]\oplus
[U_{D}(\Lambda ,a) \otimes U_{\pi}(\Lambda ,a)]\oplus
\]
\[
[U_{D}(\Lambda ,a) \otimes U_{\pi'}(\Lambda ,a)]\otimes
[U_{N}(\Lambda ,a) \otimes U_{N'}(\Lambda ,a)\otimes U_{\pi}(\Lambda ,a)]\oplus
\]
\[
[U_{N}(\Lambda ,a) \otimes U_{N'}(\Lambda ,a)\otimes U_{\pi'}(\Lambda ,a)]\oplus
\]
\beq
[U_{D}(\Lambda ,a) \otimes U_{\pi}(\Lambda ,a)\otimes U_{\pi'}(\Lambda ,a)].
\label{a.11b}
\eeq

%These is no-zero
%particle subspace because we are only considering the baryon number 2
%subspace.

%In the three-body subspaces there is not enough energy to produce
%additional pions - so we expect the two-body interactions in the
%three-body sector to be related to two-body interactions at the lower
%energy scale by cluster properties.  New three-body interactions can
%appear in the three-body sectors.  The two-body interaction in the
%two-body sector are also new because they involve energies above the
%pion-production threshold.

%The challenge is to include all of these interactions in a manner that 
%preserves both the Poincar\'e symmetry and cluster properties.

\section{Dynamics}
\label{WP_sec:4} 

In a quantum mechanical model a Poincar\'e symmetry is implemented 
by a unitary representation of the Poincar\'e group\cite{wigner}.
The representation is necessarily dynamical\cite{dirac}.  

The free-particle representation, $U_0(\Lambda, a)$, is reducible on
the model Hilbert space ${\cal H}$.  Poincar\'e group Clebsch-Gordan
coefficients \cite{moussa}\cite{coester}\cite{bkwp} can be used to
decompose the Hilbert space into a direct integral of invariant
subspaces on which $U_0(\Lambda,a)$ acts irreducibly.
The
Clebsch-Gordan coefficients have the form
\[
\langle (12)\vert 3 \rangle = 
\]
\beq
\langle (m_1,j_1)\mathbf{p}_1, \mu_1 (m_1,j_2)\mathbf{p}_2, \mu_2 
\vert (\mathbf{k}^2 , j_3 ) \mathbf{p}_3, \mu_3, d \rangle
\label{a.12} 
\eeq
where $\mathbf{k}^2$ is a more convenient label for the 
invariant mass $m_3$ of the combined system  
\beq
m_3 = \sqrt{m_1^2 + \mathbf{k}^2} + \sqrt{m_2^2 + \mathbf{k}^2} .
\eeq
In what follows we use $m_3$ and $\mathbf{k}^2=k^2$ interchangeably. 
Multiple copies of invariant subspaces with the same mass and spin
are separated by invariant degeneracy parameters $d$. 
The parameters $d$ are related
to the squares of spin $s^2$ and orbital angular momentum $l^2$ of the 
$12$-pair.
Three particle irreducible representations can be constructed by
successive pairwise coupling.

The irreducible basis states are labeled by the same
quantities that are used to label the single-particle states; linear momentum,
mass, spin, spin projection, as well as a number of Poincar\'e
invariant degeneracy parameters which we label by $d_n$, where 
$n \in \{ NN$,$NN\pi$,$NN\pi'$,$D\pi$,$D\pi'$,$D\pi\pi \}$ .  
These eigenstates are complete on the model Hilbert space.  A basis of 
non-interacting Poincar\'e irreducible states consists of the 
generalized vectors
\beq
\vert (m,j)\mathbf{p},\mu, d_n,n \rangle  = 
\left ( 
\begin{array}{c}
0 \\
\vdots \\
\vert (m,j) \mathbf{p},\mu, d_n \rangle \\
\vdots \\
0 \\
\end{array} 
\right ) . 
\label{a.12a}
\eeq 
The non-interacting invariant mass operator, $M_0$, is the mass 
Casimir operator of $U_0(\Lambda,a)$.  It
has a continuous spectrum
and is a multiplication operator in the representation (\ref{a.12a}). 
Interactions of the form \cite{bakamjian}:
\[
\langle (m',j')\mathbf{p}',\mu', d_{n'}',n' \vert V 
\vert (m,j)\mathbf{p},\mu, d_{n},n \rangle 
= 
\]
\beq
\delta (\mathbf{p}'-\mathbf{p})\delta_{j'j}\delta_{\mu'\mu} 
\langle  m', d_{n'}',n' \Vert V^j 
\Vert m, d_n,n \rangle 
\label{a.13}
\eeq
are added to the non-interacting mass operator, $M_0$,  to construct a 
dynamical mass operator:
\beq
M= M_0+V .
\label{a.14}
\eeq
Simultaneous eigenstates $\vert (\lambda,j) \mathbf{p}, \mu , d 
\rangle$ 
of $M,j,\mathbf{p},\mu$ are complete.  The dynamical problem is to 
solve the eigenvalue problem
\beq
(M_0+V) \vert (\lambda,j) \mathbf{p}, \mu , d \rangle =
\lambda \vert (\lambda,j) \mathbf{p}, \mu , d \rangle
\label{a.14a}
\eeq
in the non-interacting irreducible basis.  The eigenfunctions in the
non-interacting irreducible representation can be expressed in terms
of wave functions $\Psi_{\lambda,j,d} (m',d_{n'}', n')$:
\[
\langle (m',j') \mathbf{p}', \mu' ,d_{n'}',n' \vert (\lambda,j)
\mathbf{p}, \mu , d \rangle =
\]
\beq
\delta (\mathbf{p}'-\mathbf{p})\delta_{j'j}\delta_{\mu' \mu}
\Psi_{\lambda,j,d} (m',d_{n'}', n'). 
\label{a.15}
\eeq
There is a natural dynamical unitary representation of the Poincar\'e 
group defined on these eigenstates by
\[
\langle (m',j') \mathbf{p}', \mu' ,d_{n'}',n'  \vert U(\Lambda ,a)
\vert  (\lambda,j) \mathbf{p}, \mu , d 
\rangle =
\]
\beq
\Psi_{\lambda,j,d} (m',d_{n'}' n') 
{\cal D}^{\lambda j}_{\mathbf{p}, \mu; \mathbf{p}', \mu'} [\Lambda ,a] 
\label{a.16}
\eeq
where $\lambda$ the eigenvalue of $M$.  Completeness of the
eigenstates ensures that this representation is defined on any state.
The dynamics enters through the appearance of the mass eigenvalue,
$\lambda$, in the Poincar\'e group Wigner function ${\cal D}^{\lambda
  j}_{\mathbf{p}, \mu; \mathbf{p}', \mu'} [\Lambda ,a]$.

This construction can be done for any finite number of degree of
freedom system.  What is relevant in this construction is that the
spin $j$ in the dynamical model is the same as the spin $j$ in the
non-interacting model.  We refer to this construction of dynamical
representation of the Poincar\'e group as the generalized
Bakamjian-Thomas\cite{bakamjian} construction. 

Three different two-body models are needed as input to the 
full dynamical model.  These two-body models are a $\pi-N$  model,
a $\pi-\pi$ model and a coupled channel $NN\leftrightarrow D\pi$ 
model.  In what follows we replace the invariant mass $m_3$ by the 
relative momentum variable $k=\sqrt{\mathbf{k}^2}$.  For the coupled channel model
the non-interacting invariant mass operator has the form 
\beq
M_0 :=
\left (
\begin{array}{cc}
2\sqrt{{k}_{NN}^2+m_N^2}& 0 \\
0 & \sqrt{{k}_{D\pi}^2+m_D^2} +\sqrt{{k}_{D\pi}^2+m_\pi^2}
\end{array}
\right ).
\label{b.1}
\eeq
The interaction $V$ is defined by the kernel in the irreducible two-body 
variables  
\[
\delta (\mathbf{p}'-\mathbf{p})
\delta_{j'j}\delta_{\mu' \mu} \times 
\]
\beq
\left (
\begin{array}{cc}
\langle k_{nn}', l', s' \vert v_{NN;NN}^j \vert k_{nn}, l, s \rangle &
\langle k_{nn}', l', s' \vert v_{NN;D\pi}^j \vert k_{D\pi}, l, 1 \rangle \\
\langle k_{D\pi}', l', 1 \vert v_{D\pi;NN}^j \vert k_{nn}, l, s \rangle &
\langle k_{D\pi}', l', 1 \vert v_{D\pi;D\pi}^j \vert k_{D\pi}, l, 1 \rangle 
\end{array} 
\right ) .
\label{b.2} 
\eeq
The dynamical mass operator on 
${\cal H}_{NN} \oplus {\cal H}_{D\pi}$ is $M=M_0+V$, where
$V$ is the interaction defined by the kernel (\ref{b.2}).
Simultaneous eigenstate of $M$, $j$, $\mathbf{p}$, and
$\hat{\mathbf{z}} \cdot \mathbf{j}$ from a complete set of basis 
vectors that transform irreducibly with respect to a dynamical 
representation $U_{NN-D\pi} (\Lambda ,a) $ 
\[
\langle (\lambda',j') \mathbf{p}', \mu',d' \vert U_{NN-D\pi} (\Lambda ,a)
\vert (\lambda,j) \mathbf{p}, \mu,d \rangle = 
\]
\beq
\delta_{j'j} \delta (\lambda' -\lambda) \delta_{d'd}
{\cal D}^{\lambda j}_{\mathbf{p}, \mu; \mathbf{p}', \mu'}
[\Lambda ,a]
\label{b.3} 
\eeq 
where $\lambda$ is the eigenvalue of the dynamical mass operator $M$.
A similar construction can be done for the $\pi-\pi$ and $\pi-N$ systems,
resulting in the representations $U_{NN-\pi D} (\Lambda ,a)$ on ${\cal
  H}_{NN} \oplus {\cal H}_{D\pi}$, $U_{N-\pi} (\Lambda ,a)$ on ${\cal
  H}_{N\pi}$ and $U_{\pi\pi} (\Lambda ,a)$ on ${\cal H}_{\pi\pi}$.
These models are only required to fit scattering below the
pion-production threshold.  Because the bound states are treated as
physical degrees of freedom, the spectrum of the mass operator in
these two-body models is continuous.

\section{Algebraic cluster properties}
\label{WP_sec:5} 

Cluster properties dictate how two-body interaction are embedded in
the three-body sectors of the model Hilbert space.  Cluster properties
are normally implemented by translations.  In this framework
translations used to separate subsystems only operate on subspaces of
the Hilbert space were the translations make sense; for example it
makes no sense to separate nucleons on a subspace containing the
deuteron.  Translations that asymptotically separate subsystems ensure
that the short-ranged interactions between particles in separated
clusters vanish, however the difficulty in Poincar\'e invariant quantum
mechanics is that the translations can cause interactions that should
not vanish to vanish in the cluster limit.  This {\it will not happen
  if result of simply turning-off interactions results in a tensor
  product of subsystem unitary representations} of the Poincar\'e
group.  We call this type of clustering algebraic clustering
and use it in what follows.

Consider the problem of nucleon-nucleon scattering for invariant
energies between the one and two pion-production threshold.  The model
Hilbert has six orthogonal sectors with different particle contents.
There are three three-body sectors, each having at least one pion.  If
we translate one of the particles away from an interacting pair in one
of the three-body sectors, then the invariant energy remaining for the
interacting pair is insufficient to produce another pion.  Algebraic
cluster properties requires that in this limit the Poincar\'e
generators in this sector should become the sum of one-body
generators corresponding to the spectator particle and dynamical
two-body generators associated with an interacting two-body systems
having insufficient invariant energy to create an additional pion.
This is equivalent to the dynamical unitary representation of the 
Poincar\'e group becoming a tensor product.
In
each of these sectors there are three different two-body interactions
that arise from cluster properties, depending on which particle is
asymptotically separated.

These are the only constraints that cluster properties place on this
model.  The two-body sectors are relevant for invariant energies 
above the threshold for the production of a pion and in principle do not
have to be related to the lower-energy two-body Hamiltonians. 

The problem of two interacting particles and a spectator can be 
solved using two different methods.  The first one is to construct the 
two-body unitary representation of the Poincar\'e group,
$U_{NN - D\pi}(\Lambda,a)$, $U_{N\pi}(\Lambda,a)$,
$U_{\pi\pi'}(\Lambda,a)$ and then take the tensor product with the 
spectator representation
\beq
U_{NN - D\pi}(\Lambda,a)\otimes U_{\pi'}(\Lambda,a), \, 
\label{b.4}
\eeq
\beq
U_{N\pi}(\Lambda,a)\otimes U_{N'}(\Lambda,a), \,
\label{b.5}
\eeq
\beq
U_{\pi\pi'}(\Lambda,a)\otimes U_{D}(\Lambda,a) .
\label{b.6} 
\eeq
The second method uses the standard construction where we first use 
Poincar\'e Clebsch Gordan coefficients to decompose the non-interacting 
three-body representations into a direct integral of non-interacting irreducible 
representations.  Interactions of the from (\ref{a.13}) are added to the
non-interacting mass, which is then diagonalized to construct $U(\Lambda,a)$.
We refer to these two representations as the tensor product representation
and the Bakamjian-Thomas representation respectively.
I denote the Bakamjian-Thomas representations by 
\beq
U_{NN - D\pi;\pi'}(\Lambda,a),  
\label{b.7}
\eeq
\beq
U_{N\pi;N'}(\Lambda,a),
\label{b.8}
\eeq
\beq
U_{\pi\pi';D}(\Lambda,a)  .
\label{b.9} 
\eeq
These operators do not act on the full Hilbert space.  They act on a
subspace of the full Hilbert space.  They can be extended to operators
on the full Hilbert space by extending them to be the identity on the
orthogonal complement of the subspace.

The kernel of the two-body interaction in the two-body kinematic irreducible
representation (\ref{a.13}) depends on the kinematic two-body relative
momentum and kinematically-invariant two-body degeneracy parameters.
By coupling irreducible representations in the appropriate order, the
same variables also appear in the three-body kinematic irreducible basis.  Using
the same kernel, with different delta functions, it is easy to show
that the tensor product and Bakamjian-Thomas representations of the
dynamics of two interacting particles and a spectator give the same
$2+1$ S-matrix.  This follows because the delta functions that
multiply the interaction also appear in the $S$-matrix, and they
become equivalent when the $S$-matrix is evaluated on-shell.  While
the $2+1$ unitary representations of the Poincar\'e group are
not-equivalent, a theorem of Ekstein \cite{ekstein} implies that each
representation is related by a unitary transformation, $W$, that also
preserves the $S$ matrix.

Thus these two representations are related by
\[
W_{NN-D\pi;\pi'}  U_{NN-D\pi;\pi'} (\Lambda,a)   
W^{\dagger}_{NN-D\pi;\pi'} =
\]
\beq 
U_{NN-D\pi} (\Lambda,a) \otimes U_{\pi'} (\lambda ,a) 
\eeq
\beq
W_{N\pi;N'}  U_{N\pi;N'} (\Lambda,a)   
W^{\dagger}_{N\pi;N'} = U_{N\pi} (\Lambda,a) \otimes U_{N'} (\Lambda ,a) 
\eeq
\beq
W_{\pi\pi';D}  U_{\pi\pi';D} (\Lambda,a)   
W^{\dagger}_{\pi\pi';D} = U_{\pi \pi'} (\Lambda,a) \otimes U_{D} (\Lambda ,a).
\eeq
Because the interactions in the Bakamjian-Thomas representation 
commute with the kinematic spin, they can be combined with the 
remaining short-ranged interactions in a manner
that leads to an overall interaction of the form (\ref{a.13}).
The Bakamjian Thomas construction then be used to construct a
dynamical representation of the Poincar\'e group on the model 
Hilbert space.

The unitary operators $W_x$ operate on a subspace of the Hilbert space.
They can be extended to unitary operators on ${\cal H}$ by setting 
them equal to the identity on the orthogonal complement of the 
space on which they are defined.

\section{Construction of the dynamical mass operator} 
\label{WP_sec:6} 

The dynamical mass operator for our model in the Bakamjian-Thomas
representation has three distinct types of contributions

The first is the mass operator $M_0$ for the non-interacting system 
{\footnotesize
%\begin{center}
%{\bf Structure of $M$}
%\end{center}
\beq
M_0 = \left (
\begin{array}{cccccc}
M_{0NN'} &0&0&0&0&0\\
0 & M_{0D\pi} &0&0&0&0\\
0 & 0 & M_{0D\pi'} &0&0&0\\
0 & 0 & 0 & M_{0NN'\pi}&0&0 \\
0 & 0 & 0 & 0 & M_{0NN'\pi'}&0 \\
0 & 0 & 0 & 0 & 0 & M_{0D\pi\pi'} 
\end{array} 
\right )
\label{a.18}
\eeq
}
where the non-zero entries in this matrix are the invariant masses
of the systems of particles on each subspace.  This is the limiting 
form of this mass operator when all interactions are switched off.

Next we add lower-energy two-body interactions in each of the 
three-particle sectors:
{\footnotesize
\beq
V_2= 
\left (
\begin{array}{cccccc}
0&0&0&0&0&0\\
0&0&0&0&0&0\\
0&0&0&0&0&0\\
0 & 0 & 0 &v_{NN'} + v_{N\pi} + v_{N'\pi} & 0 & v_{NN;\pi D} \\
0 & 0 & 0 & 0 & v_{NN'} + v_{N\pi'} + v_{N'\pi'} & v_{NN;\pi' D} \\
0 & 0 & 0 & v_{NN;\pi D}^{\dagger} & v^{\dagger}_{NN;\pi' D} & v_{D\pi} + v_{D\pi'} + v_{\pi\pi'} 
\end{array} 
\right ).
\label{a.19}
\eeq
}

These are the interactions that come from the two-body 
Bakamjian-Thomas representations.  They are related to the 
tensor product representations by the unitary transformations
$W$.  For example: 
{\footnotesize
\[ 
\left (
\begin{array}{cccccc}
0&0&0&0&0&0\\
0&0&0&0&0&0\\
0&0&0&0&0&0\\
0 & 0 & 0 &v_{NN'} & 0 & v_{NN;\pi D} \\
0 & 0 & 0 & 0 & 0 & 0 \\
0 & 0 & 0 & v_{NN;\pi D}^{\dagger} & 0 & v_{D\pi} 
\end{array} 
\right ) =
\]
}
{\footnotesize
\[ 
W^{\dagger}_{NN\leftrightarrow D\pi'} 
\left (
\begin{array}{cccccc}
0&0&0&0&0&0\\
0&0&0&0&0&0\\
0&0&0&0&0&0\\
0 & 0 & 0 &m_{NN'\otimes \pi} & 0 & m_{NN;\pi' D\otimes \pi} \\
0 & 0 & 0 & 0 & 0 & 0 \\
0 & 0 & 0 & m_{NN;\pi' D\otimes \pi}^{\dagger} & 0 & m_{D\pi';\otimes \pi} 
\end{array} 
\right )
W^{\dagger}_{NN\leftrightarrow D\pi'}  -
\]
}
{\footnotesize
\beq
\left (
\begin{array}{cccccc}
0&0&0&0&0&0\\
0&0&0&0&0&0\\
0&0&0&0&0&0\\
0 & 0 & 0 &m_{0NN' \pi} & 0 & 0 \\
0 & 0 & 0 & 0 & 0 & 0 \\
0 & 0 & 0 & 0 & 0 & m_{0D\pi'\pi} 
\end{array} 
\right ).
\label{a.20}
\eeq
}
The third class of operators, $V_3$ are fully connected operators.
These are two-body interactions in the two-body sectors, three-body
interactions in the three-body sectors, $2-2$ interactions coupling
different two-body sectors and $2-3$-body interactions coupling two
and three-body sectors.  All of the two-body interactions
in $V_3$ describe
physics above the pion-production threshold and thus do not have to be
related to the interactions below the pion-production threshold.  In
addition to being connected, all of these interactions must have
kernels of the form (\ref{a.13}) in the non-interacting Poincar\'e
irreducible representation.  These connected operators do not
contribute to the cluster limit associated with lower energy
subsystems.

In this way the operator $M_{BT} = M_0 + V_2 + V_3$ commutes with 
$j$ and commutes with and is independent of $\mathbf{p}$ and $
\mathbf{j}\cdot \hat{\mathbf{z}}$.  Simultaneous eigenstates of these 
operators are complete and transform irreducibly as mass $\lambda =$ eigenvalue 
of $M_{BT}$, spin $j$, irreducible representations of the Poincar\'e group.  
We denote the resulting unitary representation of the Poincar\'e group by 
$U_{BT} (\Lambda ,a)$.  When interactions between cluster $b$ and $c$ 
in one of the three-body sectors are turned off, $U_{BT} (\Lambda ,a)$
fails to break up into a tensor product.  Instead it becomes:
\beq
U_{BT} (\Lambda ,a) \to W_{bc} [ (U_{b}(\Lambda ,a) \otimes 
U_{c}(\Lambda ,a))\oplus \cdots ] W^{\dagger}_{bc}  
\eeq

This can be repaired defining a symmetric product of the 
$W$ operators
\beq
W := e^{\sum \ln W_a} 
\eeq
where the sum runs over 
\[
a \in \{ [(NN-D\pi); \pi'],
[(NN-D\pi'); \pi],  (\pi \pi';D),(D\pi; \pi'),
\]
\beq
(D\pi';\pi),
(N\pi;N'),(N'\pi;N),(N\pi';N'),(N'\pi';N)\}.
\eeq
This operator is designed so 
\[
U (\Lambda ,a):= 
\]
\beq
W^{\dagger} U_{BT} (\Lambda ,a)W 
\to  [ (U_{b}(\Lambda ,a) \otimes 
U_{c}(\Lambda ,a))\oplus \cdots ]   
\eeq
when the interactions between particles in cluster $b$ and $c$
are turned off.  This shows that it satisfies algebraic cluster 
properties.  Because $W$ is unitary,
$U (\Lambda ,a):= W^{\dagger} U_{BT} (\Lambda ,a)W$,  is also a dynamical 
representation of the Poincar\'e group.

The completes the construction of the dynamics for the two-nucleon system
of energies between the one and two-pion production threshold. 
Thus we have a two-particle model that describes nucleon-nucleon 
scattering below the one-pion-production threshold, and we have a second model 
that described nucleon-nucleon scattering for invariant energies between the one
and two-pion-production thresholds.  
While the models are uncoupled,  interactions from the lower energy 
two-body model 
appear in the three-body sector of the higher-energy model.
Both models are exactly Poincar\'e invariant with interactions that 
can be directly constrained by few-body experiments. 
 
This general construction can be repeated inductively at higher energy 
scales.  By using the doublets the entire inductive construction used
in the fixed number of particle case \cite{fcwp82} can be generalized 
to this setting.

\section{Outlook}

This work demonstrates that it is possible to overcome all of the
difficulties in needed to extend the successful low-energy few-body
program to higher energy scales.  In this formalism the independent
degrees of freedom are taken as physical particles.  This also applies
to bound states.  This is not a problem in principle, but it does mean
that much of the work done in constructing realistic few-body
interactions needs to be repeated in this setting.  Another new
feature is that the many-body interactions of the non-relativistic
theory are replaced by both many-body interactions and hard
(high-energy) two-body interactions.  At each successive energy scale
it is necessary to introduce new hard two-body interactions.  In most
cases these interactions appear in the entrance channel and thus
cannot be ignored.  An independent method for constructing these hard
interaction would prove to be very useful in this framework.  Still, a
good part of the high-energy dynamics is strongly constrained by the
few-body dynamics through cluster properties.

This work was performed in part under the auspices of the
U. S. Department of Energy, Office of Nuclear Physics, contract
No. DE-FG02-86ER40286 with the University of Iowa.

\section{Bibliography}
\label{WP_sec:8}

References should be cited in the text by \verb+\cite{ }+
\LaTeX-commands. They should be numbered in the order in which they are
cited. See also the following examples:

%\bibitem{SchmidtPL_RefJ}
%A.A.~Author, Journal \textbf{Volume}, (year) page numbers
%\bibitem{SchmidtPL_RefB}
%B.B.~Author, \textit{Book title} (Publisher, place year) page numbers
%\bibitem{SchmidtPL_RefNP} 
%N. Surname1 and N. Surname2, Nucl. Phys. \textbf{Xnum},  
%(year) page numbers 
%\bibitem{SchmidtPL_RefPR} 
%N.N. Surname3 \textit{et al.}, Phys. Rev. \textbf{Xnum} (year) page 
%\bibitem{SchmidtPL_RefPL} 
%N. Surname4, N.N. Surname5, and N. Surname6, Phys. Lett.  
%\textbf{Xnum} (year) page numbers 
%\bibitem{SchmidtPL_RefProc}
%N.N. Surname7, \textit{Proc. of the TitleOfProceedings} (eds. NameOfEditors,  
%place, year) page numbers. 
% etc

\end{document}